\magnification=1200
\baselineskip=20 pt

\def\rc{r_c}
\def\etamunud{\eta_{\mu\nu}}
\def\dmul{\partial_{\mu}}
\def\dnul{\partial_{\nu}}

\def\dmuu{\partial^{\mu}}
\def\l{\lambda}
\def\etamunuu{\eta^{\mu\nu}}
\def\vphi{\langle \phi \rangle}
\def\hphi{\hat {\phi}}
\def\mphi{m_{\phi}}
\def\a{\alpha}
\def\b{\beta}
\def\mh{m_H}
\def\g{\gamma}
\def\d{\delta}
\def\k{\kappa}
\def\L{\Lambda}

\centerline{\bf Quantum effects of compactified {\bf AdS$_5$ }
geometry on the
higgs potential}
\vskip .5 true in
\centerline{\bf Uma Mahanta}
\centerline{\bf Mehta Research Institute}
\centerline{\bf Chhatnag Road, Jhusi}
\centerline{\bf Allahabad-211019, India}
\vskip .4 true in
\centerline{\bf Abstract}

In this paper we determine the one loop radiative correction to the 
higgs potential due to quantum fluctuations about the background metric of 
Randall-Sundrum model. We then examine the effects of the one loop
effective potential on the stability of the classical vacuum paying 
particular attention to the tadpole terms which could dominate over
the classical potential for small field configurations. We find that
although the one loop potential due to scale fluctuations could develop
a tadpole term in certain regions of the parameter space, it is positive 
for either sign of H. The quadratic and quartic terms in the radiative
correction are too small to cause any instability to the classical vacuum
for field configurations that do not violate the limits of 
ordinary perturbation theory.

\vfill\eject

\centerline{\bf Introduction}

Recently several proposals based on theories in extra dimensions have been
put forward to explain the hierarchy problem. Among them the Randall-Sundrum
model is particularly interesting since it proposes a five dimensional
world based on a non-factorizable metric

$$ds^2= e^{-2k\rc | \theta |}\etamunud dx^{\mu} dx^{\nu}-\rc^2 d\theta^2
.\eqno(1)$$

Here $\rc$ measures the size of the extra dimensions which is an ${S^1
\over Z_2}$ orbifold. $\theta$ is the coordinate of the extra dimension
with $\theta$ and $-\theta$ identified. k is a mass parameter of the order
of the fundamental Planck mass M. Two 3 branes are placed at the orbifold
fixed points $\theta =0$ (hidden brane) and $\theta =\pi$ (visible
brane). Randall and Sundrum showed that any field on the visible brane
with a fundamental 
 mass parameter $m_0$ gets an effective mass $m=m_0 e^{-k\rc\pi}$
due to the exponential warp factor. Therefore
 for $k\rc \approx 12$ the weak scale
is generated from the Planck scale by geometry. Note that the 
compactification scale $\mu_c \approx {1\over \rc}\approx {k\over 12}$
that is  required to generate the large hierarchy is only an order of 
magnitude smaller than M unlike in theories with large extra dimensions.

In the Randall-Sundrum model $\rc$ is the vacuum expectation value of a 
massless scalar field T(x). The modulus is therefore not stabilized
by some dynamics. In order to stabilize the modulus Goldberger and Wise 
introduced a scalar field $\chi(x, \theta )$ in the bulk with interaction
potentials localised on the branes. This they showed could generate a
potential for $T(x)$ and stabilize the modulus at the right value 
($k\rc\approx 12$) needed for the hierarchy without any excessive
fine tuning of the parameters.

The couplings of the radion (scale fluctuations) and the graviton (tensor
fluctuations) to the SM higgs scalar localized on the visible brane is
completely determined by general covariance. In this paper we shall
determine the couplings of the radion and the graviton to the SM higgs
scalar. We then use these couplings to derive the one loop effective
potential of the higgs boson arising from fluctuations about
the background metric. Finally  we examine the effect of the radiative
corrections on the stability of the classical vacuum by paying particular
attention towards the possible occurrence of tadpole terms in the radiative
corrections which could dominate over the classical potential for small
H.

\centerline{\bf Radion couplings to the higgs scalar}

We shall determine the effects of scale fluctuations and tensor 
fluctuations on the higgs potential separately. First consider the
effect of scale fluctuations. The couplings of the radion to the higgs 
scalar can be determined from the following action
$$S=\int d^4x \sqrt {-g_v}[g_v^{\mu\nu}{1\over 2}\dmul H \dnul H -V(H)].
\eqno(2)$$

where $V(H)={1\over 2} m_H^2 H^2+\l v H^3+{\l\over 4}H^4$. H is a small
fluctuation of the higgs field from its classical vacuum v. The constant
vacuum energy $V(0)$ has been subtracted out from $V(H)$. In the
abscence of scale fluctuations we have
$g_v^{\mu\nu}= e^{2k\pi T(x)}\eta^{\mu\nu}=({\phi\over f})^{-2}\etamunuu$,
$\sqrt{-g_v} =({\phi\over f})^4 $ and $\phi = f e^{-k\pi T(x)}$. Here
f is a mass parameter of the order of the fundamental Planck mass M.
Rescaling H
 and v as $H\rightarrow {f\over \vphi}h$ and $v\rightarrow
{f\over \vphi} v$ we get

$$S=\int d^4x [({\phi\over \vphi})^2{1\over 2}\etamunuu\dmul H \dnul H-
({\phi\over \vphi})^4 V(H)].\eqno(3)$$

Let $\hphi$ denote a small fluctuation of the radion from its vev
$\vphi$. We then get

$$S=\int d^4x[ {1\over 2} \dmul H \dmuu H -V(H)]
+\int d^4x[\dmul H \dmuu H -4 V(H)]{\hphi\over \vphi}+...\eqno(4)$$

where the dots represent terms of order $({\hphi\over \vphi})^2$ and higher.
It is easy to see that the couplings of
 cubic and higher order fluctuations to the higgs field
do not contribute to the one loop effective potential. Further in this 
paper to simplify matters
 we shall ignore the effects of the couplings of
 quadratic radion fluctuations to the higgs scalar.
 Under this approximation we need to consider 
only vertices with one and only one radion field. A more complete treatment
that takes into account the effects of the couplings of
 quadratic radion fluctuations into account will be given
elsewhere.

\vfill\eject

\centerline{\bf Radion contribution to the one loop effective potential}

We would like to note that since in the evaluation of the effective 
potential the external lines are assumed to have zero momenta the kinetic 
energy term of the higgs scalar in eqn. (4) does not contribute 
to the one loop effective potential. To simplify our discussion we shall
first consider the radion coupling only to the mass term of the
higgs scalar. The generalization to the case where the radion couplings to
all the three terms of V(H) are taken into account is straightforward
and will be presented at the end of this section. We shall follow the
method of Coleman-Weinberg and sum up an infinite series of one loop 
diagrams with an arbitrary number of higgs lines at zero momenta
attached to it. From each $\hphi hh$ vertex we have to choose one 
external  higgs line leaving one $\hphi$ and one higgs line to go into
the loop. It therefore follows that only loop daiagrams with even number
of external higgs lines need to be considered. Summing all such one loop
diagrams we get

$$\eqalignno{\delta V&=\sum_{n=1}^{\infty}V_{2n}\cr
&=-{(m_H^2)^2\over 16\pi^2}\int x dx [{1\over 2}{\beta^2\over y^2}
+{1\over 4}{\beta^4\over y^4}+..+{1\over 2n}{\beta^{2n}\over y^{2n}}..]\cr
&={(m_H^2)^2\over 32\pi^2}\int x dx [\ln (y^2-\beta^2)-\ln y^2].&(5)\cr}$$

where $\alpha = {\mphi^2\over m_H^2}$, $y^2=(x+\alpha )(x+1)$
and $\beta ={4H\over \vphi}$.

Evaluating the above integral with a physical cut off $M_s$ (the string
scale) we get

$$\eqalignno{ \delta V_r &=V_0(M_s)+{(\mh^2)^2\over 32\pi^2}[({M^2_s\over
\mh^2})^2 (\ln {M_s^2\over \mh^2}-{1\over 2})+(1+\a )({M_s^2\over \mh^2})\cr
&-{1\over 2}(1+\a^2)({1\over 2}+\ln{M_s^2\over \mh^2})]-{(\mh^2)^2\over
32\pi^2}\b^2\ln{M_s^2\over \mu^2}+{(\mh^2)^2\over 32\pi^2}[\b^2
(\ln{\mh^2\over \mu^2}-{1\over 2})\cr
&+{\g^2\over 2}\ln \g+{\d^2\over 2}\ln \d.] &(6)\cr}$$

where $V_0(M_s)=-{(\mh^2)^2\over 32\pi^2}\int x dx \ln (x+\a )(x+1) dx$,
$\g ={(1+\a )\over 2}-{1\over 2}\sqrt{(1-\a )^2+4\b^2}$
and $\d ={(1+\a )\over 2}+{1\over 2}\sqrt{(1-\a )^2+4\b^2}$.
$\mu$ is the renormalization or subtraction scale. The terms that
are independent of H contributes to the vacuum energy and could be
subtracted out. The term $\beta^2 \ln{M_S^2\over \mu^2}$
can be absorbed into the mass counter term of the higgs scalar.
The remaining H dependent terms are finite (independent of the cut
off $M_s$)
 and they represent
the renormalized one loop correction to the higgs potential
due to linear radion fluctuations. In deriving the above result
we have considered the radion coupling only to the mass term of
V(H). If we consider the radion couplings to all the three terms
of V(H) then each vertex can be either $\hphi H^2$, $\hphi H^3$
or $\hphi H^4$. It can be shown that under this condition the
radiative correction (unrenormalized) to the higgs potential
is given by

$$\d V={(\mh^2)^2\over 32\pi^2}\int x [\ln(y^2-(\b+\b^{\prime}+\b^{\prime
\prime})^2)-\ln(x+\a )(x+1)]dx.\eqno(7)$$

where $\b^{\prime}={6H^2\over v\vphi}$ and $\b^{\prime\prime}={2H^3\over
v^2\vphi}$. We find that for small fluctuations of the higgs field from
 its classical vacuum (i.e. for $H<v$) the contributions of $\hphi H^3$
and $\hphi H^4$ vertices to the higgs potential are suppressed compared
to that of $\hphi H^2$. 

\centerline{\bf Graviton couplings to the higgs scalar}

We shall now consider the effects of graviton fluctuations on the
higgs potential. In the following discussion we shall ignore the radion
fluctuations and take the metric to be

$$ds^2= e^{-2k\rc |\theta |} g_{\mu\nu} dx^{\mu}dx^{\nu}-\rc^2 d\theta^2.
\eqno(8)$$

where $g_{\mu\nu}=\etamunud +\kappa h_{\mu\nu}$.

The graviton couplings to the higgs scalar are completely determined
by general covariance and can be derived from the action given in
eqn (2).

As in the case of the radion we shall be interested only in the 
radiative corrections that arise from the couplings of linear graviton
fluctuations to the higgs field. Under this approximation we have

$$S=\int d^4x ({\vphi\over f})^4 (1+{\kappa\over 2}h_{\a}^{\a})
[{1\over 2} ({\vphi\over f})^2 (\etamunuu -\kappa h^{\mu\nu})
\dmul H\dnul H -V(H)].\eqno(9)$$

Rescaling the higgs field and its vev properly we get

$$\eqalignno{S &=\int d^4x[{1\over 2}\etamunuu \dmul H\dnul H-V(H)]\cr
&+\int d^4x [{1\over 4}h^{\a}_{\a}\etamunuu \dmul H\dnul H -{1\over 2}
h^{\a}_{\a} V(H)-{1\over 2}h^{\mu\nu}\dmul H\dnul H].&(10)\cr}$$

The graviton couplings to the higgs scalar is therefore given by

$$\eqalignno{S& =-{\k\over 2}\int d^4x h^{\mu\nu}(x, \pi) T_{\mu\nu}(H)\cr
&=-{\k\over 2}\sum_{n=0}^{\infty}\int d^4x h_n^{\mu\nu}(x){\chi _n(\pi)
\over \sqrt {\rc}}T_{\mu\nu}(H).&(11)\cr}$$

where $T_{\mu\nu}(H)=\dmul H\dnul H-\etamunud [{1\over 2}
\partial_\a H\partial^a H-V(H)]$.
It can be shown that as long as the mass $m_n$ of the Kaluza-Klein
mode is much small compared to the fundamental Planck mass M,
 $\chi_n(\pi)\approx e^{k\rc\pi}
\sqrt{k\rc}$ and $\chi_0(\pi)=\sqrt{k\rc}$. Choosing $\k ={2\over M_p
^{3\over 2}}$ we finally get

$$S=-{1\over M_p}\int d^4x h_0^{\mu\nu}(x) T_{\mu\nu}(H)
-{1\over \Lambda}\sum_{n=1}^{\infty}\int d^4x h_n^{\mu\nu}(x)
T_{\mu\nu}(H)\eqno(12)$$

where $\Lambda \approx M_p e^{-k\rc\pi}\approx O(Tev)$.

So the coupling of the zero mode to the higgs scalar (or any other
 SM field) is inversely
proportional to the Planck scale but the coupling of the higher
Kaluza-Klein modes to the higgs scalar is inversely proportional to the
Tev scale. It is this feature that makes the Kaluza-Klein modes 
accessible at Tev energies. As in the case of the radion the K.E. term
of the higgs field does not contribute to the one loop effective potential
which is computed with the external lines at zero momentum.
Also the contributions of the cubic and quartic terms in V(H) can
be neglected compared to that of the mass term of the higgs scalar.
So the relevant interaction Lagrangian for us is given by
$L_I=-{m_H^2\over 2\Lambda}\sum_{n=1}^{\infty} h_n^{\mu\nu}\etamunud H^2$.

Using this interaction it can be shown that

$$\eqalignno{V_2 &={1\over 2}\sum_n ({m^2_H H^2\over \L})^2
\int {d^4k\over (2\pi )^4} D^n_{\mu\nu\alpha\beta}(k){1\over (k^2-m^2_H)}
\eta^{\mu\nu}\eta^{\alpha\beta}\cr
&= -\sum_n ({m^2_H H^2\over \L})^2 \int {d^4k\over (2\pi )^4} {2\over 3}
{({k^2\over m_n^2}-2)({k^2\over m_n^2}+1)\over (k^2+m_n^2)(k^2+m_H^2)}\cr
&=-{(m^2_H )^2\over 16 \pi^2}\sum_n {1\over 2} ({a_nH\over \L})^2
\int_0^{M^2_s\over m_H^2} x dx [{2\over 3}f(x)]^2.&(13)\cr}$$

$$V_4=-{(m^2_H)^2\over 16\pi^2}\sum_n {1\over 4}({a_nH\over \L})^4
\int_0^{M^2_s\over m_H^2} x dx [{2\over 3}f(x)]^4$$

and 

$$V_{2m}=-{(m^2_H)^2\over 16\pi^2}\sum_n {1\over 2m} 
({a_nH\over \L})^{2m} [{2\over 3}f(x)]^{2m}$$

where $a_n={m^2_H\over m_n^2}$.
and $f(x)=\sqrt {(x-{2\over a_n})
\over (x+1)}$.
The mass $m_n$ of the Kaluza-Klein
modes are given by $m_n=e^{-k\pi r_c}k x_n$ where $x_n$ are the roots
of the Bessel function $J_1(x_n )$ [6]. 
 The sum runs over all the massive Kaluza-Klein graviton
modes. Adding all such contributions arising from diagrams with
an even number of external higgs lines we get

$$\eqalignno{\delta V_g &= {(m^2_H)^2\over 32\pi^2}\sum_n\int x dx \ln (1-
{a_n^2\over \L^2}H^2 f^2(x))\cr
&=[V_0(M_s)+{M_s^4\over 64\pi^2} (\ln{M_s^2\over m_H^2}-1)]+
{(m^2_H)^2\over 32\pi^2}[{1\over 2}{M_s^4\over m_H^4}\ln (1-\a_n)\cr
&+{(1+2{\a_n \over a_n})\over (1-\a_n)}{M_s^2\over \mphi^2}
-{(1+2{\a_n\over a_n})^2\over 2(1-\a_n)^n}\ln {M_s^2\over \mu^2}]
+{(m^2_H)^2\over 64\pi^2}[{(1+2{\a_n\over a_n})^2\over (1-\a_n)^2}
(\ln{M_h^2\over \mu^2}-{1\over 2})\cr
&-{(1+2{\a_n\over a_n})\over (1-\a_n)^2}\ln (1-\a_n)
+{(1+2{\a_n\over a_n})^2\over (1-\a_n)^2}\ln (1+2{\a_n\over a_n})]
.&(14)\cr}$$

where $\a_n={2\over 3}a_n^2 {H^2\over \L^2}$ and $V_0(M_s)=-{(m^2_H)^2
\over 32\pi^2}\sum_n\int x \ln (1+x) dx$.
The mass $m_n$ of the first few Kaluza-Klein modes usually lie in the 
several Tev range. Thus if the higgs  boson is light  i.e. $m_H<$1 Tev
 the coefficients $a_n$ are much small compared to unity. Hence
for small field configurations the radiative correction due to radion
dominates over that of gravitons.
The terms inside the first pair of brackets are independent of H and
can be absorbed in the vacuum energy. The terms within the second
pair of brackets depend both on the cut off and H. These terms can be
absorbed in the counter terms that has to be added to the
classical higgs potential to make the one loop effective potential
insensitive to the cut off. The terms inside the third pair of brackets
depend on H but not on the cut off. They represent the renormalized
radiative correction to the higgs potential due to graviton fluctuations.
We would like to note that if $\a_n$ and ${\a_n\over a_n}$ are small
then $\ln (1-\a_n)$, ${(1+2{\a_n\over a_n})\over (1-\a_n)}$ and
${(1+2{\a_n\over a_n})^2\over (1-\a_n)^2}$ can be expanded into an 
infinite power series in H. Since the interaction of the KK graviton modes
to the higgs scalar is non-renormalizable we need an infinite number
of local counter terms to absorb the cut off dependent terms
of the one loop effective potential.

\centerline{\bf Effect of the radiative correction on the vacuum stability}

Examining the one loop radiative correction due to graviton fluctuations
we find that there is no potentially dangerous
tadpole term that could destabilize the classical vacuum. The lowest order
term is quadractic in the higgs field. Further it is suppressed by a
loop factor of ${1\over 64\pi^2}$ and another factor of ${m^4_H\over 
m_n^2\L^2}$ arising from KK graviton couplings to higgs scalar. The 
quadratic term arising from radiative corrections due to gravitons
is therefore smaller than the quadratic term present in the classical 
potential. Further although there are higher order terms in the
graviton induced
radiative correction these terms are small within the limits of validity
of perturbation theory. The higgs field configurations that are needed 
for these terms to dominate and cause any instability occur outside
the limits of ordinary perturbation theory.

The form of the lowest order term in the radiative correction due to the
radion is difficult to determine in general since $\gamma$ and 
and $\d$ have a complicated dependence on h. We shall therefore study it under
 two special cases.

{\bf Case I}: In this case we shall assume that $1-\a \ll 2\beta$. It then 
follows that $\g \approx {1+\a\over 2}-\b$ and 
$\d \approx {1+\a\over 2}+\b$. Substituting the values of $\g$ and
$\d$ we get

$$\eqalignno{\d V_r&\approx {(\mh^2)^2\over 32\pi^2}[\b^2 (\ln {\mh^2\over
\mu^2}-{1\over 2}+{(1+\a )\b \over 2}\ln {{1+\a\over 2}+\b\over 
{1+\a\over 2}-\b}\cr
& +{(1+\a)^2\over 8}(1+{4\beta^2\over (1+\a)^2})\ln ({(1+\a)^2\over 4}-\b^2)]
.&(15)\cr}$$

Therefore in this case the radion induced radative correction give a 
tadpole term 
 which can dominate over the quadaratic and quartic terms 
in the classical potenitial for small H. However the tadpole term is 
positive for both signs of h and therefore it
 does not destabilize the classical vacuum.

{\bf Case II}: We shall now consider the case when $1-\a \gg 2\b$. It then 
follows that $\g\approx \a-{\b^2\over (1-\a)}$
and $\d\approx 1+{\b^2\over (1-\a)}$. Substituting the values of
$\g$ and $\d$ in $\d V_r$ we get

$$\eqalignno{\d V_r&\approx {(\mh^2)^2\over 32\pi^2}[\b ^2(\ln {\mh^2\over
\mu^2}-{1\over 2})+{1\over 2}(\a -{\b^2\over 1-\a})^2\ln (\a-{\b^2\over
1-\a})\cr
&+{1\over 2}{\b^2\over (1-\a)}(1+{\b^2\over 1-\a})^2].&(16)\cr}$$

The lowest order term 
 in this case is quadartic in the higgs field. However
 it is suppressed by a loop factor of ${1\over 32\pi^2}$ and an
additional factor
of ${\mh^2\over \vphi^2}$ coming from radion couplings.
Hence in this case we do not expect any insatability to the classical
vacuum to develop within the limits of perturbation theory.

\centerline{\bf Conclusion}

In this paper we have determined the one loop radiative correction to the 
higgs potentail due to linear radion and graviton fluctuations in the 
Randall-Sundrum model. We find that the radiative correction due to
graviton fluctuations do not give rise to tadpole term. On the other hand
although the radiative correction due to radion fluctautions does give rise
to a tadpole term under some cases it is positive for both sign of H.
The quadratic and higher order terms in the radiative correction 
for both kinds of fluctuations
are too small compared to the classical potential for any instability
to develop within the limits of ordinary perturbation theory.

\centerline{\bf References}

\item{1.} L. Randall and R. Sundrum, Phys. Rev. Lett. 83, 3370 (1999).

\item{2.} N. Arkani-Hamed, S. Dimopoulos and G. Dvali, Phys. Lett.
B 429, 263 (1998) and Phys. Rev. D 59, 086004 (1999).

\item{3.} W. D. Goldberger and M. B. Wise, Phys. Rev. Lett. 83, 4922 
(1999).

\item{4.} W. D. Goldberger and M. B. Wise, Phys. Lett. B 475, 275 (2000).

\item{5.}S. Coleman and E. Weinberg, Phys. Rev. D 7, 1888 (1973).

\item{6.} W. D. Goldberger and M. B. Wise, Phys. Rev. D 60, 107505 (1999);
H. Davoudiasl, J. L. Hewett and T. G. Rizzo, Phys. Rev. Lett. 84,
2080 (2000).

\end